\documentclass[a4paper,10pt,twoside]{cpc-hepnp}

\usepackage{multicol}
\usepackage{graphicx}
\usepackage{booktabs}
\usepackage{amssymb,bm,mathrsfs,bbm,amscd}
\usepackage[tbtags]{amsmath}
\usepackage{lastpage}

\begin{document}
%\begin{CJK*}{GBK}{song}
\fancyhead[co]{\footnotesize ...
{ }} %\footnotetext[0]{Received }

\title{Study of BESIII electromagnetic calorimeter performance with radiative lepton pair events \thanks{ Supported in part
by the CAS/SAFEA International Partnership Program for Creative
Research Teams, CAS and IHEP grants for the Thousand/Hundred Talent programs and National Natural Science Foundation of China under Contracts No. 11175189.}}

\author{%
\quad Vindhyawasini ~Prasad$^{1,2;1}$\email{vindy@ihep.ac.cn}%
\quad Chunxiu ~Liu$^{1}$ %\email{xinchou@ihep.ac.cn}%
\quad Xiaobin ~Ji $^{1}$ %\email{xinchou@ihep.ac.cn}%
\quad Weidong ~Li$^{1}$ %\email{xinchou@ihep.ac.cn}%
\quad Huaimin ~Liu$^{1}$ %\email{xinchou@ihep.ac.cn}%
\quad Xinchou ~Lou$^{1,3}$ %\email{xinchou@ihep.ac.cn}%
}
\maketitle

\address{%
$^1$ Institute of High Energy Physics, Chinese Academy of
Sciences, Beijing 100049, China\\
$^2$ CMR Institute of Technology, Bangalore 560037, India \\
$^3$ University of Texas at Dallas, Richardson, Texas 75080-3021, USA\\

}

\begin{abstract}
We study the photon detection efficiency and position resolution of the electromagnetic calorimeter (EMC) of the BESIII detector. The control samples of the initial-state-radiation (ISR) process of $e^+e^-\rightarrow \gamma \mu^+\mu^-$ at $J/\psi$ and $\psi(3770)$ resonances are used  for the calibration of the photon  cluster shapes and photon detection efficiency study.  The photon detection efficiency is defined as the fraction of predicted photon, determined by performing a kinematic fit with the four momenta of two charged tracks only, matched with the actual photons in the EMC. The spatial resolution of the EMC is studied in polar ($\theta$) and azimuthal ($\phi$) angle directions in a cylindrical coordinate system centered at the interaction point, with z-axis along the beam direction. 
\end{abstract}



\begin{multicols}{2}

\section{Introduction}
The BESIII experiment is a high intensity electron-positron collider experiment located at the Institute of High Energy Physics, Beijing, China \cite{EMC}. It  has collected a large amount of data at different center-mass-energies between ($2.0-4.6$) GeV, including the $J/\psi$ and $\psi(3770)$ resonances,  to study the hadron spectroscopy and search for new physics phenomena in the tau-charm region. Other low-energy lying charmonium states can be produced via a radiative photon emission process from the high-energy resonances, such as $J/\psi \rightarrow \gamma \eta_c$ decay. The masses and widths of the measured charmonium  states are limited by the resolution of the photon energy and position measurements if it is measured from the recoiling mass of the photon. BESIII uses a CsI(Tl) based electromagnetic calorimeter (EMC) to measure the electromagnetic (EM) showers of electrons and photons with excellent energy and angular resolutions \cite{bes3_nim}. The energy (position) resolution of the BESIII EMC is $2.5\%/\sqrt{E~(\rm{GeV})}$ ($6$ mm $/\sqrt{E ~({\rm GeV})}$). 

The incident particles, interact with the materials of the EMC detector, loss their energy and thus generate the EM showers \cite{EMCAL}. Typical EM shower tends  over many adjacent crystals and form a cluster of adjacent energy deposits. EMC distinguishes the charged particles against neutral as in case of the energy deposited through charged tracks should have an associated track, while the neutrals should have only energy deposits. The photons have very short and compact cluster shapes, while hadrons and other particles have very broad and scattered cluster shapes. The pattern recognition algorithms are used to analyze the shower shapes and differentiate the real photons from the fake photons  \cite{babar}. The energy of the showering particles in the EMC is defined as the energy deposited in the $3\times3$ or $5\times5$ EMC crystals \cite{He}.

One of the important tasks of the BESIII EMC is to measure the energy and position of electrons, photons and neutral particles  with excellent energy and position resolutions. Any bias in the position measurement of the showering particles in the EMC could result a systematic shift in the kinematic variables, e.g. four-momenta of neutral pions, thus degrade the position resolution of the EMC and introduce the bias in the measurements of the physical parameters. Therefore, the possible biases have to be studied carefully and must be corrected as much as possible. The position resolution of the EMC also depends on the crystal performance \cite{bes3_nim}. 

In this paper, we describe the study of photon detection efficiency and position resolution of the BESIII EMC detector. The radiative muon pair events are used for the studies of energy and position resolutions of the EMC, and photon detection efficiency. We also describe the procedure used to calibrate the cluster shapes of the photons, which can separate the real photons from the fake photons. The position resolution of the EMC is defined as the separation in polar ($\theta$) and azimuthal ($\phi$) angles  between the reconstructed and expected positions of the charged track in the front face of the EMC crystals.
\section {BESIII  detector}
\label{DetectorandDS}
The BESIII spectrometer is a multi-purpose device designed to
simultaneously measure many properties of the particles
produced in $e^+e^-$ collisions near the tau-charm region, as
described in detail in \cite{bes3_nim}. It has a geometrical acceptance of $93\%$ of $4\pi$ and has four detector sub-components. Charged particle momenta are measured in a 43 layers helium-based ($40\%$ He, $60\%$ $C_3H_8$) main drift chamber, operating in a 1.0 T solenoidal magnetic field.  A scintillation based time-of-flight system (TOF), which has one barrel and two end-caps,  and the energy loss (d$E$/d$x$) in the tracking system are used for charged particle identification. The energy of photons and electrons is measured by EMC, while muons are identified using a muon chamber (MuC), which contains nine (eight) layers of resistive plate chamber counter interleaved with steel in barrel (end-cap) region.  

\subsection {BESIII  electromagnetic calorimeter} 
BESIII EMC contains 6,240 CsI(Tl) crystals and  has one barrel and two end-cap regions \cite{bes3_nim}. The length of each crystal is 28 cm or $15.1 X_{0}$, where $X_{0}$ is the radiation length. The barrel region contains 5,280 crystals which are divided into 44 rings, denoted by $\theta_{index}$, and thus each ring contains 120 crystals. All the crystals are tilted within $1.5$ degree in $\phi$ directions and $1.5 - 3.0$ degrees in the $\theta$ direction ($\pm 5$ cm away from the interaction point in the beam direction) to avoid photons from the interaction point escaping through cracks between crystals. Each end-cap region contains 6 rings. The number of crystals in the six rings of each end-cap is 96, 96, 80, 80, 64 and 64. The end-caps contain 33 different sizes of crystals, and among the 960 crystals in the end-cap regions, 192 crystals are irregular pentagons.

\section{ Data and MC}
We study the EMC performance using $2.93$ ($0.08$) $fb^{-1}$ of the data  collected at $\psi(3770)$ ($J/\psi$) resonance  during 2009 to 2011 \cite{psipps}. The same amount of the generic $J/\psi$ and  $\psi(3770)$ decays of the events, simulated by EvtGen package  \cite{EVTGEV}, are used for the background studies. The Bhabha scattering and di-photon events are generated by BABAYAGA  \cite{babayaga} while the PHOKHARA \cite{phokhara} package is used for simulating the events for the  initial state radiation (ISR) channels of $e^+e^- \rightarrow \gamma \mu^+\mu^-$, $e^+e^- \rightarrow \gamma \pi^+\pi^-$,  $e^+e^- \rightarrow \gamma \pi^+\pi^-\pi^0$. The detector response and time dependent reconstruction efficiencies are determined by a MC simulation based GEANT4 package  \cite{GEANT4} and have been included in the simulated events.

\section{Study of EMC performance with radiative muon pair events} 
\label{sec:2}
 We use the control sample of the ISR process of $e^+e^- \rightarrow \gamma \mu^+\mu^-$ at $J/\psi$ and $\psi(3770)$ resonances for the energy and position resolutions of the EMC and photon detection efficiency studies. The event of interests is reconstructed with two oppositely charged tracks, where the two charged tracks are required to have their points of closest approach to beam-line within $\pm 10.0$ cm from the interaction point in the beam direction ($V_z$) and within $\pm 1.0$ cm in the plane perpendicular to the beam ($V_{x,y}$). The two charged tracks are also required to be in the detector acceptance region \cite{bes3_nim}. In order to improve the purity of the muons in the events, the penetration depth in MuC is required to be greater than 35 cm \cite{MUC}. We perform the kinematic fit using two charged tracks with a condition that the mass of the missing track must be zero. The $\chi^2$ from the kinematic fit is required to be less than 25. Due to the mass of the muon very close to the pion mass, the kinematics of both the decay processes are almost similar. The control sample of radiative pion pair events is also considered as the signal for this analysis. At this level, the  contribution of the other sources of backgrounds is less than $1\%$.  We use the photon estimate obtained from the kinematic fit to compute the energy and position resolutions of the EMC and photon detection efficiency. Figure~\ref{fig:predphot} shows the energy and cosine of polar angle distributions of predicted photon for both $J/\psi$ and $\psi(3770)$ data-sets.

\begin{center}
\includegraphics[width=8.0cm,height=8.0cm]{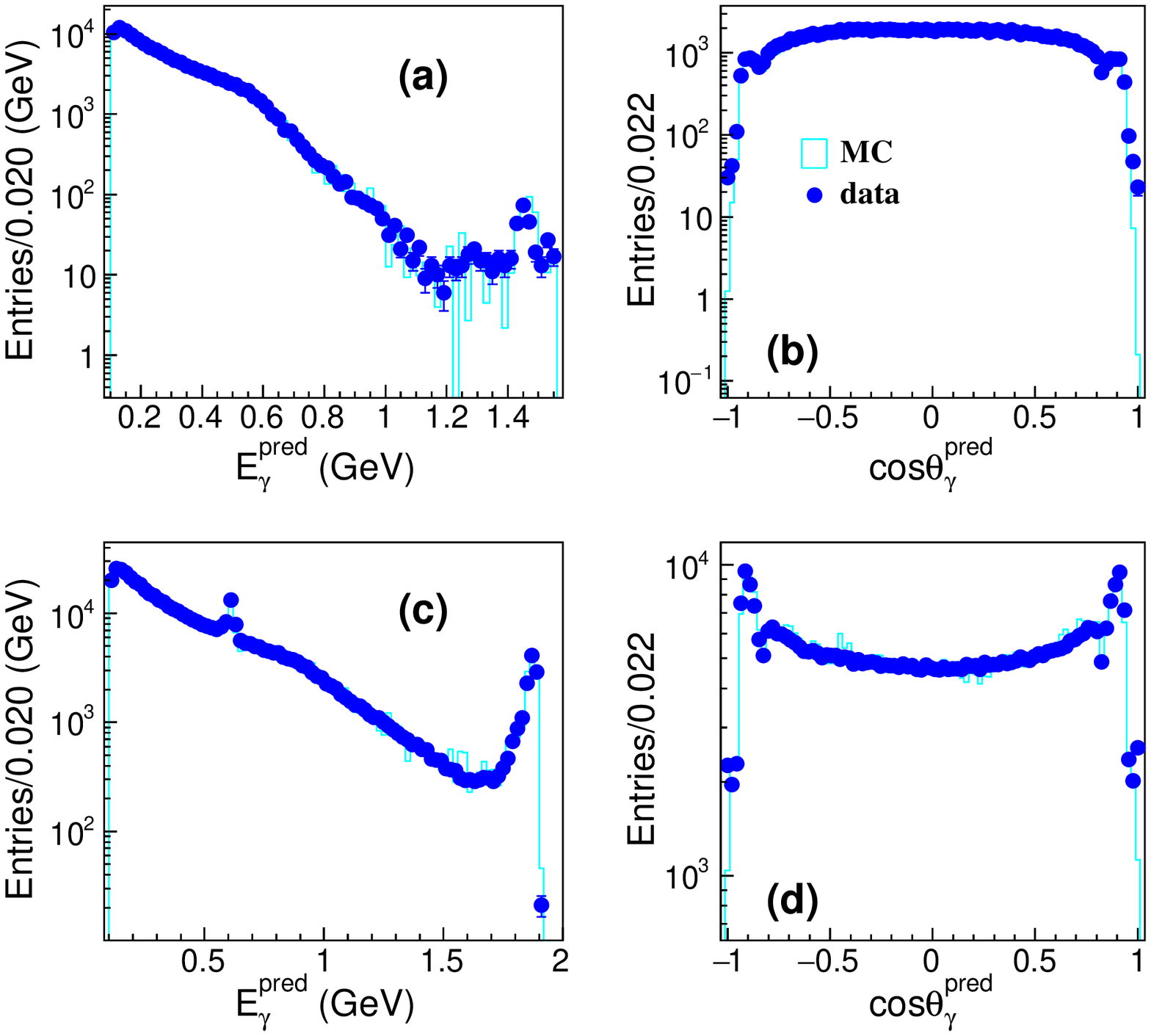}
\figcaption{\label{fig:predphot} (a,c) The energy and (b,d) cosine of polar angle distributions of predicted photon obtained from the 1C kinematic fit. The top plots are for $J/\psi$ data-set and bottom plots for $\psi(3770)$ data-set. A peak around 0.6 GeV appears in the energy distribution for $\psi(3770)$ data-set due to the ISR production of the $J/\psi$ resonance. The MCs seem to describe well with the data. }
\end{center}

\subsection{Calibration of photon cluster shape}
 The shower processes in the EMC are required to be well characterized for the purposes of particle identification and reconstruction. A number of variables have been developed to study the shower shapes of the different particles in the EMC, such as second moment and lateral moment ($LAT$)  \cite{latmom}. These two variables are used to quantify the transverse shower shape of the cluster and separate the EM showers from the hadronic showers. The EM showers tend to deposit a large fraction of their energy in one or two crystals, whereas the hadronic showers tend to be more spread out.

 The second moment is defined as $\sum_{i} E_i r_i^2/\sum_{i} E_i$, where $E_i$ is the deposited energy in the $i^{th}$ crystal and  $r_i$ the radial distance of the crystal $i$ from the cluster center. The $LAT$  \cite{latmom} is defined as,
\begin{equation}
LAT = \frac{\sum_{i=3}^{N} E_i r_i^2}{\sum_{i=3}^{N} E_i r_i^2+E_1r^2+E_2r^2}
\end{equation}
\noindent where $r$ is average distance between two crystals, which is equal to 5 cm. 

We calibrate the shapes of $LAT$ and second moment distributions while adjusting EMC incoherent noise (EINC) value \cite{EINC}. The clean sample of radiative muon pair events has been used for this study. A clear discrepancy between data and old MC, simulated with the EINC value of 0.20 MeV as  previously used by BESIII experiment, can be seen in Figure~\ref{fig:EINC}, with the data constantly shifted to the higher value of lateral and second moments. The discrepancy between data and MC has been overcome while simulating a new MC sample with the EINC value of 0.27 MeV. The $LAT$ and second moment distributions in the new MC seem to describe well with the data (Figure~\ref{fig:EINC}).

\begin{center}
\includegraphics[width=8.0cm,height=4.0cm]{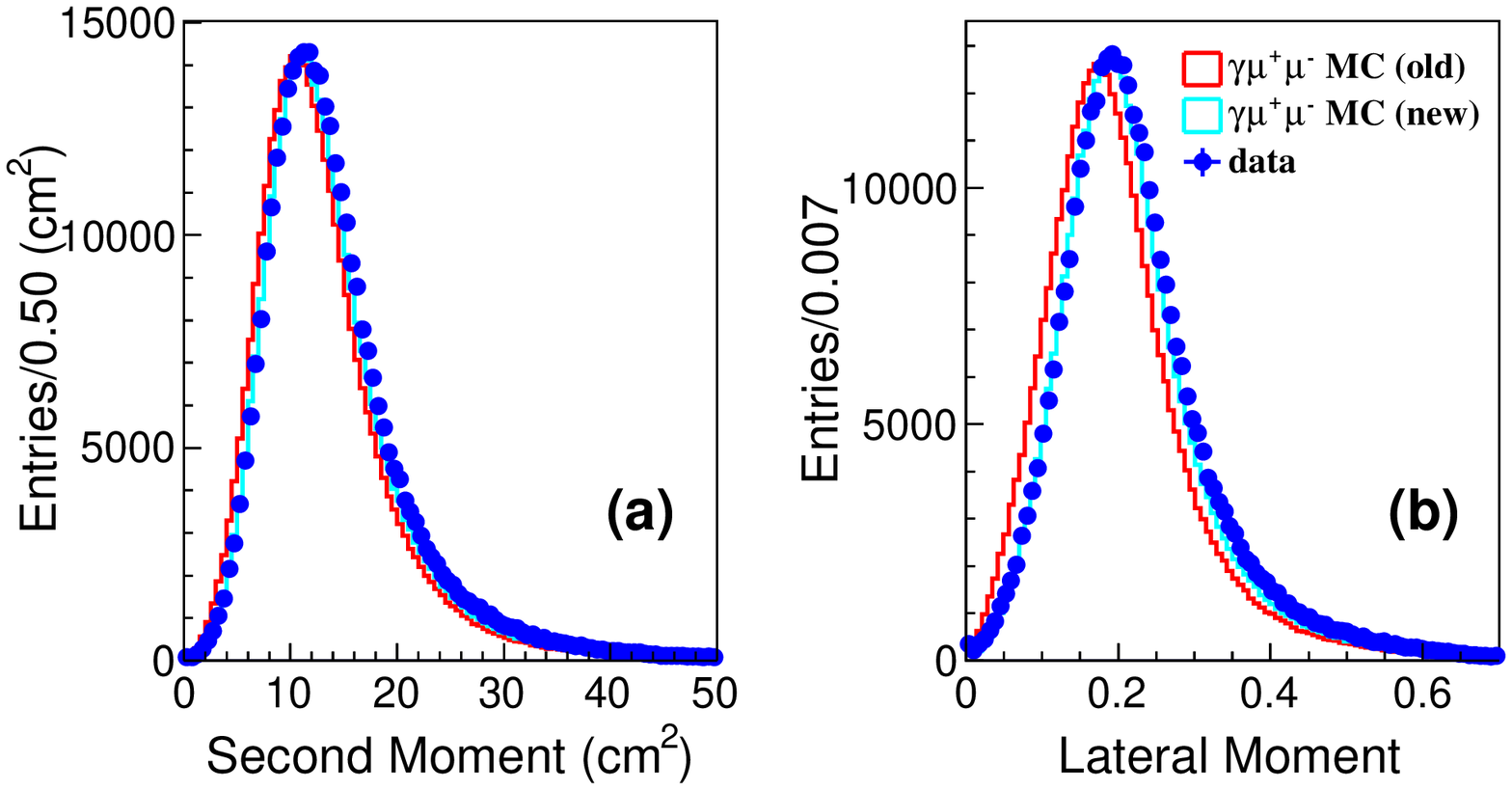}
\figcaption{\label{fig:EINC} The plots of (a) second moment and (b) lateral moment distributions for data (blue), old MC  (red) and new MC (cyan).  The old (new) MC includes the EINC value of 0.20 (0.27)  MeV. The distributions of second and  $LAT$  moments in the new MC describe well with the data.
 }
\end{center}

We study the effect of EINC in the energy and position resolutions, computed from the difference in energy and angular distributions  between predicted and reconstructed photons, at different values of predicted photon energy ($E_{\gamma}^{pred}$) using both old and new MC simulated with the EINC value of 0.20 MeV and 0.27 MeV, respectively (Figure~\ref{fig:EINCpsip_}). The energy resolution of the EMC in the new MC seems to  be a little bit worse than the older MC, but it is much closer to the real data. However, the effects of increased EINC and kinematic fit  are observed to be almost negligible in the angular distributions. The spatial resolutions are multiplied by $l$ and $r$ for $\theta$ and $\phi$ directions, respectively, to measure the unit in cm, where  $l$ is the  path length of extrapolated track between interaction point and EMC cluster centroid, and $r$ is the inner radius of the EMC. The position resolution of the EMC in old MC is compatible with new MC. The polar angle resolution seems to be worse than the expected one due to depending on the resolution of the $V_z$, which value is much larger than the resolutions of $V_x$ and $V_y$. However, due to depending on the resolutions of the $V_x$ and $V_y$, the $\phi$ resolution seems to be compatible with the expected one in the higher energy region. The energy and position resolutions, obtained from the MC-truth, seem to describe well with their values in data and MC in the high energy region. 

\begin{center}
\includegraphics[width=8.0cm,height=8.0cm]{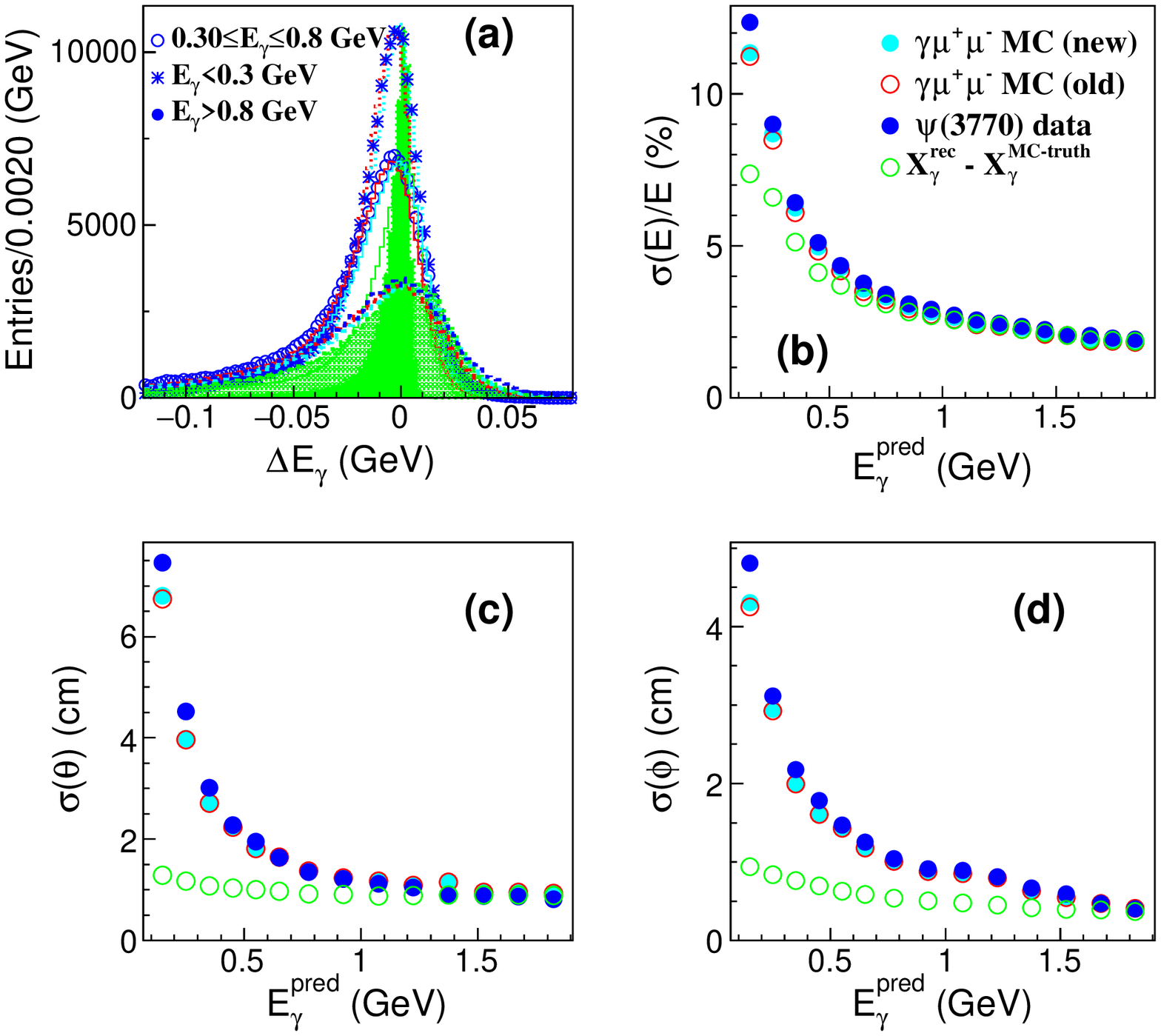}
\figcaption{\label{fig:EINCpsip_} The plots of (a) $\Delta E_{\gamma}$ distribution,  (b) energy resolution ($\sigma_{E}/E$) vs. $E_{\gamma}^{pred}$,  (c) polar angle resolution ($\sigma_{\theta}$) vs. $E_{\gamma}^{pred}$ and (d) azimuthal angle resolution ($\sigma_{\phi}$) vs. $E_{\gamma}^{pred}$  for  new MC (cyan), old MC (red) and data (blue), together with their plots obtained from the MC-truth information (green).  The  $\Delta E_{\gamma}$  is defined as the energy difference between predicted and reconstructed photons.   
 }
\end{center}

\subsection{Photon detection efficiency}
We finally compute the photon detection efficiency, which is defined as the fraction of predicted photon matched with actual photons in the EMC, using the clean sample of radiative muon pair events. If a predicted photon matched with actual photons in the EMC, the $\theta_{ \gamma,\gamma^{pred}}$ and $E_{\gamma}/E_{\gamma}^{pred}$ distributions are expected to peak at 0 radian and 1, respectively,  where $\theta_{ \gamma,\gamma^{pred}}$ is the angle between predicted and reconstructed photons,  and $E_{\gamma}/E_{\gamma}^{pred}$ the energy ratio of the reconstructed photon to the predicted photon (Figure~\ref{fig:ang}).  The regions of $\theta_{ \gamma,\gamma^{pred}} < 0.5$ radian and $E_{\gamma}/E_{\gamma}^{pred} \in [0.15, 1.4]$ are considered to be the detected region of the photon in the EMC. The photon detection efficiency as a function of energy and the cosine of polar angle for both the data and MC is shown in Figure~\ref{fig:2djpseff}.  We also compute the systematic uncertainty due to photon reconstruction, defined as the relative difference in photon detection efficiency between data and MC, to be observed up to the level of $1\%$ (Figure~\ref{fig:syst}).

\begin{center}
\includegraphics[width=8.0cm,height=4.0cm]{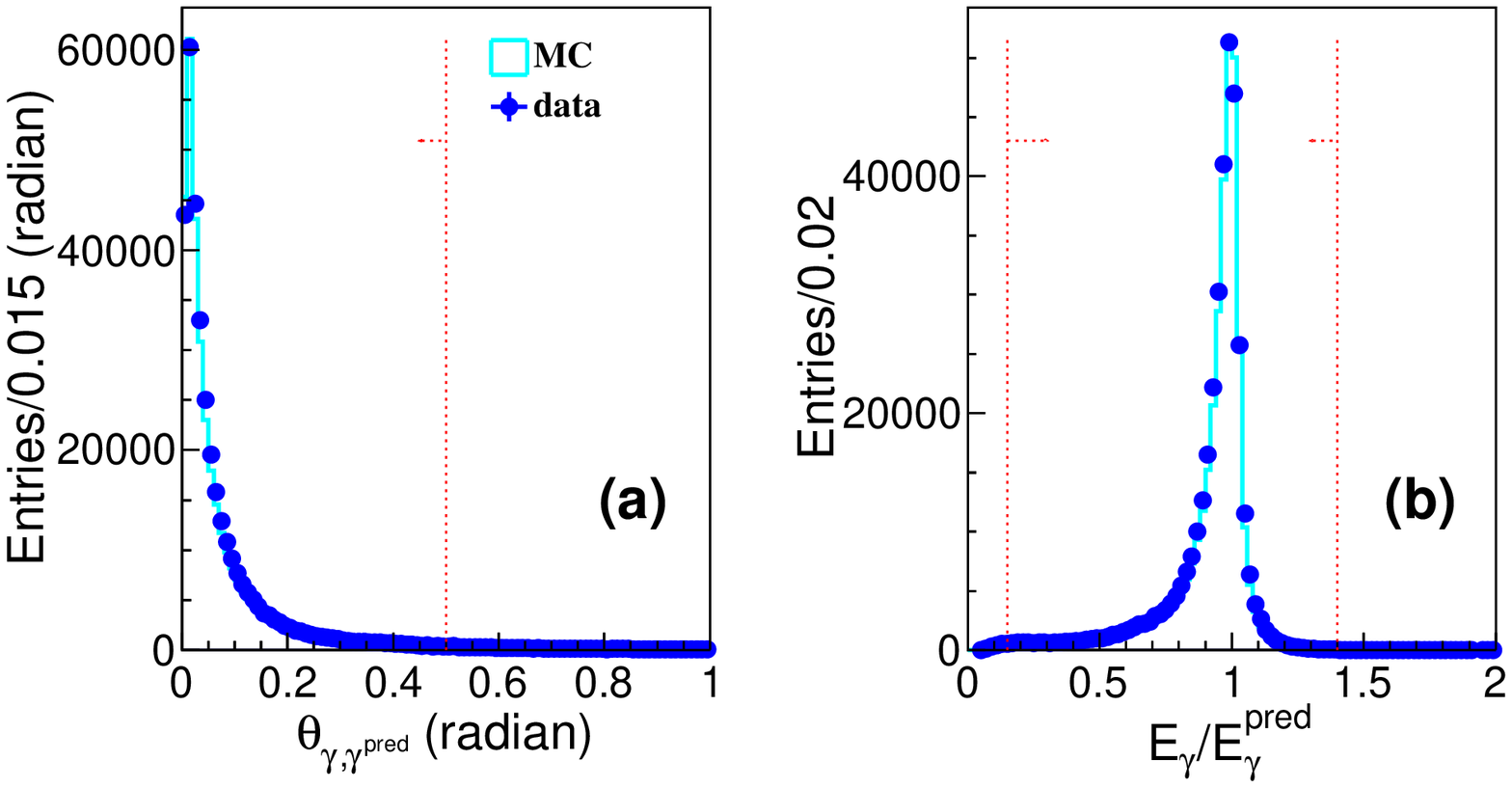}
\figcaption{\label{fig:ang} The plots of (a) $\theta_{ \gamma,\gamma^{pred}}$ and (b) $E_{\gamma}/E_{\gamma}^{pred}$ distributions for data (blue) and MC (cyan). The arrows represent the detected region of the photon in the EMC.
}
\end{center}

\begin{center}
\includegraphics[width=8.0cm,height=5.5cm]{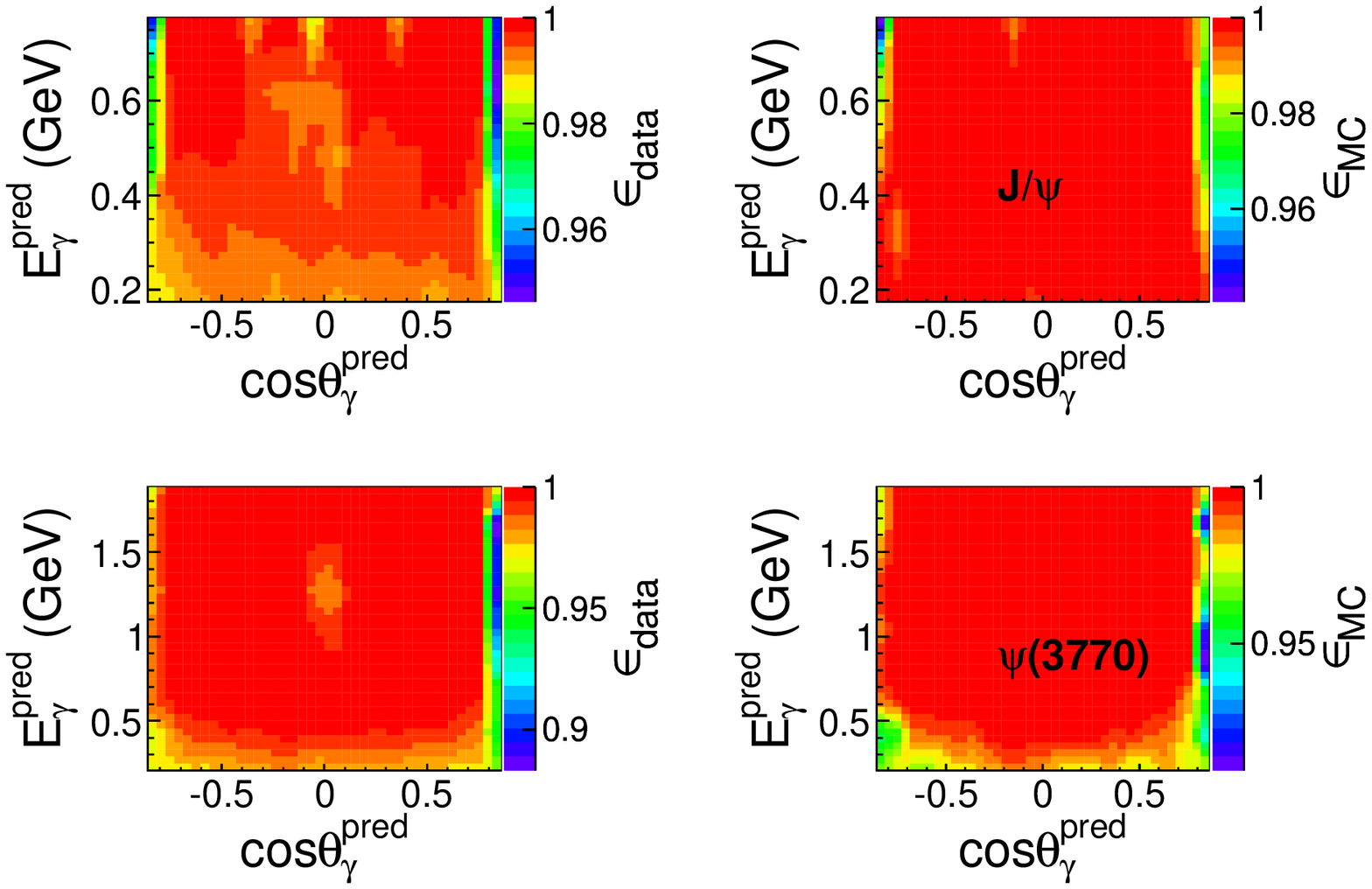}

\figcaption{\label{fig:2djpseff} The 2-d plots of the efficiency curves in $\cos\theta_{\gamma}^{pred}$ and  $E_{\gamma}^{pred}$ planes for data (left) and MC (right), where $\theta_{\gamma}^{pred}$ and $E_{\gamma}^{pred}$ are the polar angle and energy, respectively of the predicted photon obtained from the kinematic fit. Top plots are for $J/\psi$ data and bottom plots for $\psi(3770)$ data. The photon efficiency is studied only for $E_{\gamma}^{pred} \le 0.8$ GeV in the $J/\psi$ data due to low statistics as seen in Figure~\ref{fig:predphot}. 
 }
\end{center}

\begin{center}
\includegraphics[width=8.0cm,height=3.0cm]{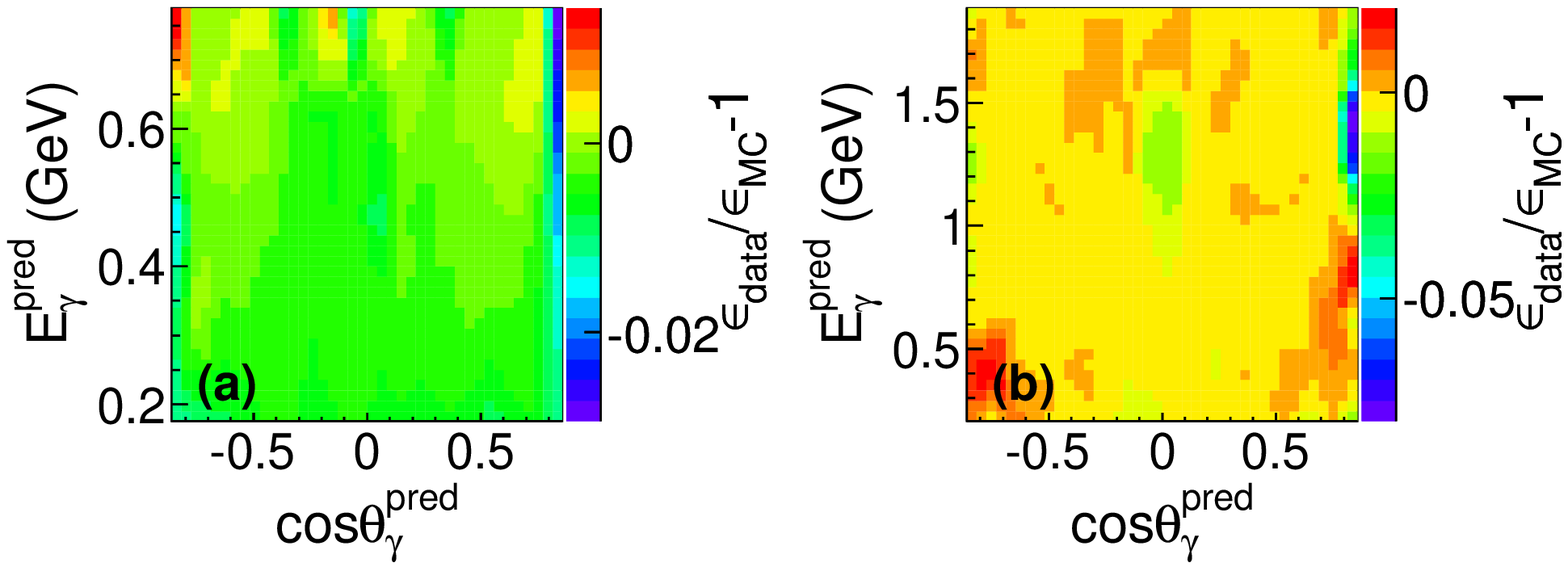}

\figcaption{\label{fig:syst} The 2-d plots of the relative difference in efficiency  between data and MC in $\cos\theta_{\gamma}^{pred}$ and  $E_{\gamma}^{pred}$ planes for (a) $J/\psi$ data  and (b) $\psi(3770)$ data, where $\theta_{\gamma}^{pred}$ and $E_{\gamma}^{pred}$ are the polar angle and energy, respectively of the predicted photon obtained from the kinematic fit. The relative differences in photon detection efficiencies between data and MC are observed up to the level of $1\%$. 
 }
\end{center}

\section{Shower position Reconstruction}

The incident particles create the EM shower inside the calorimeter. The showers develop laterally and longitudinally in the several connected crystals while losing the energy by the incident particles. The continuous connected region of the crystals deposited by energy of the incident particles is called clusters. Each shower is recognized by a seed, which is the local maxima of energy deposit
among its neighbors.  A center-of-gravity (CG) method is  used to calculate the impact coordinate, $x_c$, of the showering particles on the front face of the EMC \cite{Brabson, He}. Mathematically, the CG is defined as,  

\begin{equation}
x_c = \frac{\sum_{i=0}^n W_i(E_i)x_i}{\sum_{i}W_i(E_i)}
\label{CM}
\end{equation}

\noindent where  $W_i(E_i)$ is the weighted function of the energy in the cluster and $x_i$ the coordinates of the center of the $i^{th}$ crystals in the cluster in the front face of the EMC crystals. The sum includes the information about all crystals of the clusters. The simplified version of this method is the \rm{\lq\lq linear weighting function\rq\rq}, which is defined as,

\begin{equation}
W_i(E_i) = E_i
\end{equation}

The second approach is the \rm{\lq\lq logarithmic weighting function\rq\rq} which can reduce the weight of the most energetic crystal and enhance the low energy ones. The logarithmic weighting function is defined as, 
\begin{equation}
 W_i^{{\rm log}}(E_i) = \mathrm{Max}\{0,a_0+{\rm ln}(E_i)-{\rm ln}(E_{tot})\}
\end{equation}

 \noindent where $E_{tot}$ is the sum of total energy deposited to the crystals and $a_0=4.0$  the cutoff parameter which guarantees that the logarithm gets a positive argument and removes crystals with very low energy. BESIII experiment uses these two methods for measuring the shower positions of the particles in the EMC \cite{He}.

\subsection{Position resolution}
 We study the position resolution of the EMC using  a radiative Bhabha sample at $J/\psi$ resonance in  $\delta \theta$ and $\delta \phi$ distributions, defined as separations in ($\theta,\phi$) positions between  reconstructed (rec) and the expected position of the track on the front face of the EMC. These angular distributions are further translated as $\Delta \theta \rightarrow \delta \theta \times l$ and $\Delta \phi \rightarrow \delta \phi \times r$ to measure them in cm. Figure~\ref{performance} shows the resolution values of $\Delta \theta$ and $\Delta \phi$ distributions as a function of $e^-$ momentum. The resolution of the $\Delta \phi$ distribution degrades a little bit in the low-momentum region due to the effect of magnetic field. The position resolution of the EMC for $e^{\pm}$ seems to be compatible with the expected position resolution. Whereas in Figure~\ref{fig:EINCpsip_}, the EMC position resolution seems to be degraded due to worse resolution of the event vertex.  

\begin{center}
\includegraphics[width=8.0cm,height=4.5cm]{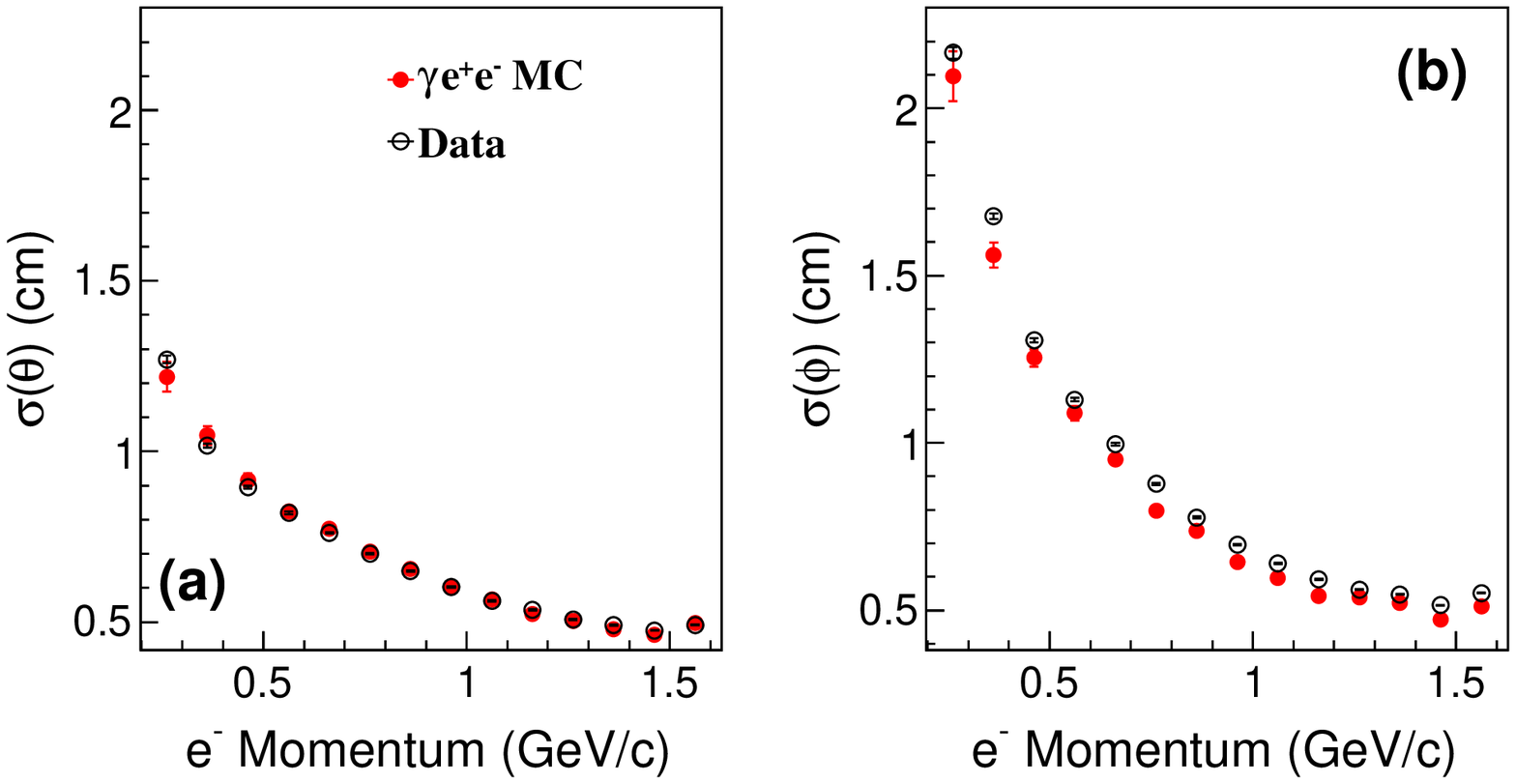}
\figcaption{\label{performance} The resolution of (a) $\Delta \theta$ and (b) $\Delta \phi$  distributions as a function of $e^-$ momentum. The resolution of $\Delta \phi$ distribution in the low-momentum region seems to be worse due to the effect of magnetic field. }
\end{center}

 We use simulated single electron and photon events to understand the discrepancy between the position resolutions for photon and charged tracks. The expected position of the tracks is computed using either the MC-truth information on the front face of the EMC crystals or the extrapolated (ext) track obtained from the MDC to determine the position resolution. The position resolution of the EMC in $\theta$ and $\phi$ directions for both electron and photons is shown in Figure~\ref{singelec}. The position resolution of the EMC in $\theta$ direction, computed using the extrapolated track information obtained from the MDC, seems to be worse due to the large resolution values of the extrapolated charged tracks. The position resolution of the EMC for both charged and neutral tracks seems to be same and compatible with the expected values in the control samples of single electron and photon events. The EMC intrinsic resolutions from photon and electron are consistent and the resolution is decided by the EMC geometry.

\begin{center}
\includegraphics[width=8.0cm,height=5.0cm]{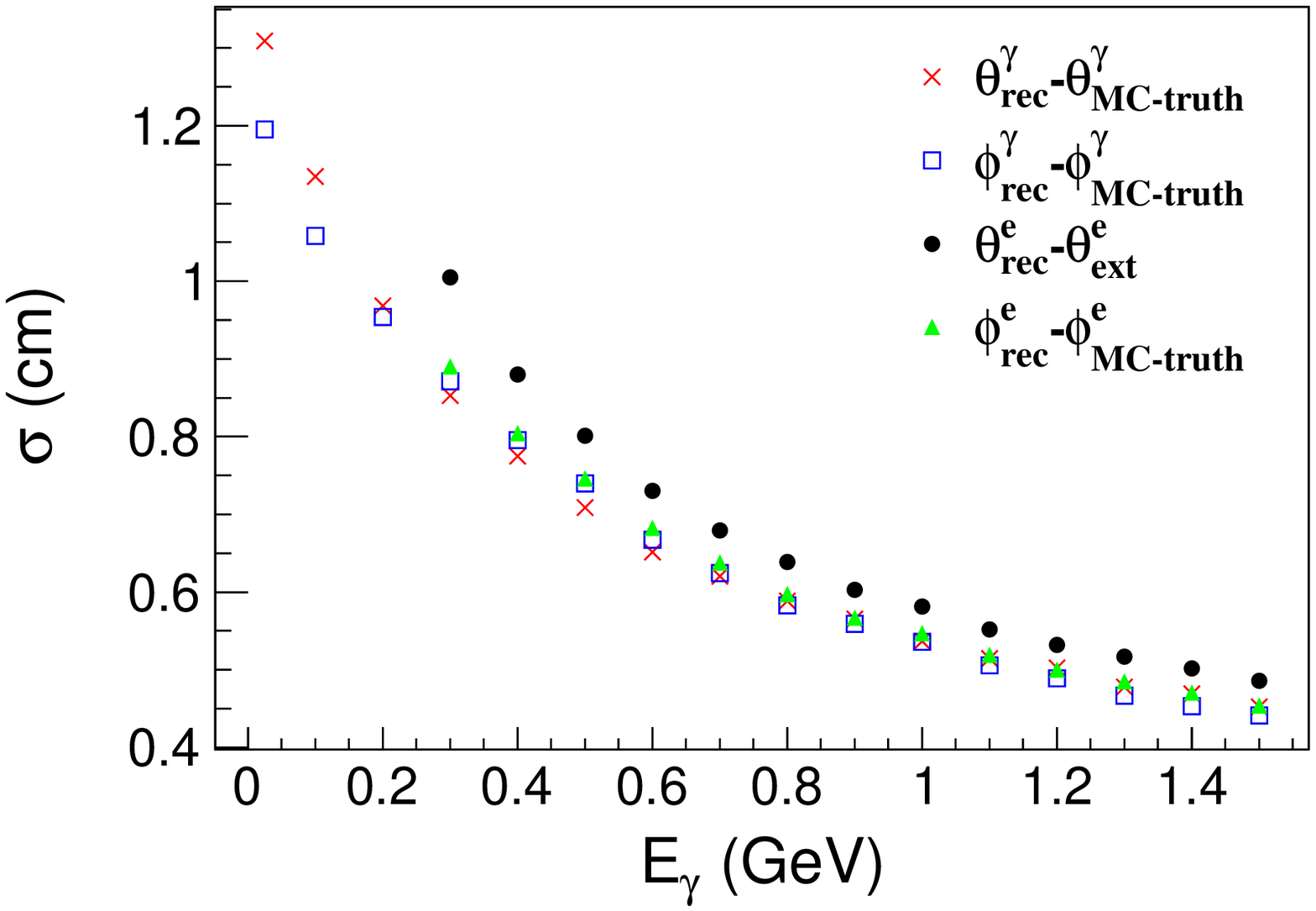}
\figcaption{\label{singelec} The position resolution of the EMC in theta and $\phi$ directions for both electron and photons computed using the MC samples of single electron and photon events. The position resolution of the EMC in $\theta$ direction, computed using the extrapolated track information obtained from the MDC, seems to be worse due to the poor resolution of the extrapolated track. The $\theta$ and $\phi$ MC-truth are the EMC MC-truth  on the front face of the crystals and they are not the MC-truth at the level of the generator.
 }
\end{center}

%%%%%%%%%%%%%%%

\section{Summary and conclusion}
We calibrate the photon cluster shapes and compute the photon detection efficiency of the BESIII detector using a clean control sample of radiative muon-pair events. The systematic uncertainty, defined as the relative difference in photon detection efficiency between data and MC, due to photon reconstruction is observed up to the level of $1\%$. The position resolution of the EMC has been studied in two independent coordinates of polar and azimuthal angles using radiative bhabha sample. The position resolution of the EMC is observed to be compatible with the expected value of the position resolution of the EMC for both charged and neutral tracks.

\end{multicols}

\clearpage
%\end{CJK*}
\end{document}